\documentclass[10pt]{article}          %%
\usepackage{mathrsfs,amsfonts}
plus.2em minus.4em \belowdisplayshortskip=2mm plus.2em minus.4em

\makeatletter

\@addtoreset{equation}{section}  
\makeatother

\title{\centerline{\normalsize}%  \hfill hep-th/0607059}
\bf{Supersymmetric
  quantum mechanical   generalized  MIC-Kepler system}}

\author{ {\bf Pulak Ranjan Giri\thanks{e-mail :pulakranjan.giri@saha.ac.in}}\\
\normalsize Saha Institute of Nuclear Physics, 1/AF Bidhan-Nagar,
Calcutta 700064, India}

\date{\today}

\begin{document}\maketitle
%%%%%%%%%%%%%%%%%%%%%%%%%
\begin{abstract} \noindent\small
We  construct  supersymmetric(SUSY) generalized  MIC-Kepler system
and show that the systems with half integral Dirac quantization
condition $\mu= \pm\frac{1}{2}, \pm\frac{3}{2},
\pm\frac{5}{2},.....$  belong to a SUSY family (hierarchy of
Hamiltonian) with same spectrum between the respective partner
Hamiltonians  except for the ground state. Similarly, the systems
with integral Dirac quantization condition $\mu =\pm 1,\pm 2, \pm
3,......$ belong to another  family. We show that, it is necessary
to introduce additional potential to MIC-Kepler system in order to
unify the two  families  into one. We also reproduce the results of
the (super-symmetric) Hydrogenic problem in our study.
\end{abstract}
%%%%%%%%%%%%%%%%%%%%%%%%%
%%%%%%%%%%%%%%%%%%%%%%%%%
\section{\small{\bf {Introduction}}} \label{in}
%%%%%%%%%%%%%%%%%%%%%%%%%
Supersymmetric (SUSY) quantum mechanics is getting more attention
and is being studied extensively nowadays. However, the concept of
SUSY came in High Energy Physics for the first time in order to
obtain all the fundamental interactions of nature
\cite{ramond,golfand,neven,volkov,wess} under a sort of unified
theory. SUSY is a highly nontrivial symmetry for fermions and
bosons. Because, fermions and bosons are particles with different
principle and statistics, for example, fermion obeys Pauli exclusion
principle but boson does not. SUSY has some predictions in High
Energy Physics. For example, every fundamental particle (i.e.,
quarks, leptons, gluon, photon, etc.) in nature must have its  SUSY
partner, whose spins differ  by half-integral but mass are equal
when the symmetry is preserved.   But so far there is no
experimental evidence that SUSY is preserved in nature. Then people
started to think about SUSY breaking. Inspired by this fact,  Witten
\cite{witten} started to study SUSY breaking in non-relativistic
quantum mechanics as a toy model for SUSY breaking in quantum field
theory. SUSY has now application in atomic, nuclear, condensed
matter and statistical physics \cite{cooper} also.

SUSY quantum mechanics has now become a separate field of study and
has received lot of interest for its beautiful mathematical insight
as well as for various aspect of non-relativistic quantum mechanics.
Non-relativistic coulomb problem is a standard problem in SUSY
quantum mechanics, for example. In Ref. \cite{alan} it has been
shown that the distinct spectrum of certain atoms and ions have
quantum mechanical SUSY. For example, the $s$ levels of the lithium
atom can be interpreted as the SUSY partner of the hydrogen atom $s$
levels in the absence of electron-electron interactions and provided
the valence electron of lithium is far enough removed from the core
electrons cloud. It is also discussed that the SUSY is broken if
electron-electron interaction is present for the obvious reason that
the level with fixed principal quantum number $n$ and different
orbital angular momentum quantum number $l$ splits up. For other
discussion on coulomb problem see Ref.
\cite{alan1,alan2,alan3,alan4}. Later it is however shown that the
relativistic Coulomb problem also have exact SUSY \cite{hosho}.

So far known knowledge of SUSY on Coulomb problem naturally raises
the question whether it can be extended to the case of two dyon
system (MIC-Kepler).  Because MIC-Kepler system \cite{zwa, mcl} is
very similar to the non-relativistic coulomb problem, where two
dyons with electric and magnetic charges $e_1, g_1$ and $e_2, g_2$
respectively are present and the radial equations of both the system
are exactly equal except the parameter orbital angular momentum. Not
only that,  MIC-Kepler system restores  all the symmetry of
non-relativistic Coulomb system like $O(4)$ for bound state and
$O(1,3)$ for continuum state if a potential of the form $\sim
\frac{1}{r^2}$ is added to the Hamiltonian by hand.  The system has
been further generalized \cite{mar,mar1} and it  is called
generalized MIC-Kepler system for obvious reason. In this present
work we will discuss about  this generalized MIC-Kepler system
\cite{mar,mar1} in the framework of SUSY quantum mechanics.

The paper is organized as follows: In Sec.~\ref{su}, we recapitulate
the basic formalism of SUSY quantum mechanics. In Sec.~\ref{mic} we
will discuss about the generalized MIC-Kepler system. Because in
order to discuss the SUSY of the model it is necessary to know about
the details of the model. Then in Sec.~\ref{susy}, the SUSY
extension of the generalized MIC-Kepler system will be discussed and
some observations will be made. Finally we conclude in
Sec.~\ref{con}.

%%%%%%%%%%%%%%%%%%%%%%%%%
\section{\small{\bf{General review of the formalism of SUSY quantum
mechanics}}}\label{su}
%%%%%%%%%%%%%%%%%%%%%%%%%%%%
In this section we recapitulate the general feature of SUSY quantum
mechanics. For detail review see
\cite{cooper,cooper3,cooper1,cooper2,cooper4} and the
references therein. The SUSY quantum mechanical Hamiltonian can be
written as %%
\begin{eqnarray}
H_{susy}(x)= H_+(x) \oplus H_-(x). \label{sysyH}
\end{eqnarray}
In $\hbar= m = 1$ unit, the two partner Hamiltonians $H_+(x)$ and
$H_-(x)$ take the form %%
\begin{eqnarray}
H_\pm(x) =-\frac{1}{2} \frac{d^2}{dx^2}+V_\pm(x)\,, \label{partner1}
\end{eqnarray}
where the SUSY  partner potentials $V_+(x)$ and $V_-(x)$ can be
written in terms of superpotential $W(x)$ as %%
\begin{eqnarray}
V_\pm(x) = W^2(x) \mp {1 \over \sqrt{2}}W^{\prime}(x)\,,
\label{partnerpotential1}
\end{eqnarray}
It can be easily checked  from Eq. (\ref{partner1}) and  Eq. (\ref
{partnerpotential1}) that the Hamiltonian $H_+(x)$ and $H_-(x)$ can
be factorized in the form %%
\begin{eqnarray}
H_+(x)= A(x)^\dagger A(x)\,,~~~ H_-(x)= A(x)A(x)^\dagger\,,
\label{factorization1}
\end{eqnarray}
if we consider the first order differential operator $A(x)$ and
$A(x)^\dagger$ of the form %%
\begin{eqnarray}
A(x) = {1 \over \sqrt{2}}{d \over dx} + W(x)~,~ A^{\dag}(x) =- {1
\over \sqrt{2}}{d \over dx} + W(x)~. \label{A}
\end{eqnarray}
We can define two supercharges $Q(x)$ and  $Q(x)^\dagger$ of the
form %%
\begin{eqnarray}
Q(x)= A(x)\sigma_-,~~~~Q^\dagger(x)= A^\dagger(x)\sigma_+,
\label{charge}
\end{eqnarray}
where $\sigma_+ $ and  $\sigma_- $ can be written in terms of Pauli
matrices as %%
\begin{eqnarray}
\sigma_\pm= \frac{1}{2} (\sigma_x \pm i\sigma_y). \label{pauli}
\end{eqnarray}
The Hamiltonian $H_{susy}(x)$ together with the two supercharge
$Q(x)$ and $Q^\dagger(x)$ form the closed super-algebra $sl(1,1)$:
\begin{eqnarray}\nonumber
[H_{susy}(x),Q(x)]& =& [H_{susy}(x),Q^{\dag}(x)] = 0 ~,
\{Q(x),Q^{\dag}(x)\} = H_{susy}(x)~~, \\   \{Q(x),Q(x)\}& =&
\{Q^{\dag}(x),Q^{\dag}(x)\}=0~. \label{sysyalgebra}
\end{eqnarray}

The SUSY partner Hamiltonians  $H_+(x)$ and $H_-(x)$, its
eigenvalues $E_n^{(+,-)}$ and the eigenfunctions are related among
themselves if it is shown that SUSY is preserved for a quantum
mechanical  system. The time independent Schr\"{o}dinger eigenvalue
equation for $H_+(x)$ takes the form %%
\begin{eqnarray}
H_+(x) \psi_n^{+}(x) = A^{\dag}(x) A(x) \psi_n^{+}(x) =
E_n^{+}\psi_n^{+}(x)\,. \label{eigen1}
\end{eqnarray}
The Hamiltonian $H_-(x)$ has also the same eigenvalue $E_n^+$ but
the eigenfunction is different.  It is easily found  by multiplying
from left on both sides of  equation (\ref{eigen1}) by the operator
$A$
\begin{eqnarray}
H_-(x) (A(x) \psi_n^{+}(x))  = A(x) A^{\dag}(x) A(x) \psi_n^{+}(x) =
E_n^{+}(A(x) \psi_n^{+}(x))~. \label{eigen2}
\end{eqnarray}
It is also possible to write the time independent Schr\"{o}dinger
eigenvalue equation for $H_-(x)$ with eigenvalue $E_n^-$ and
eigenfunction $\psi_n^-$ and from there one can show that $H_+(x)$
has also the same eigenvalue $E_n^-$ but with eigenfunction
$A^{\dag}(x)\psi_n^-(x)$.
\begin{eqnarray}
H_-(x) \psi_n^{-}(x) = A(x) A^{\dag}(x) \psi_n^{-}(x) =
E_n^{-}\psi_n^{-}(x)\,. \label{eigen12}
\end{eqnarray}
\begin{eqnarray}
H_+(x) (A^{\dag}(x) \psi_n^{-}(x))  = A^{\dag}(x) A(x) A^{\dag}(x)
\psi_n^{-}(x) = E_n^{-}(A^{\dag}(x) \psi_n^{-}(x))~. \label{eigen22}
\end{eqnarray}

It is clear that the same eigenvalue is shared by the two partner
Hamiltonians $H_+(x)$ and $H_-(x)$ but with different
eigenfunctions. The exception occurs when some eigenfunction is
annihilated  by $A(x)$. It is obvious that the eigenfunction
annihilated by the operator $A(x)$ is a ground state, but ground
state may not be annihilated by $A(x)$. The above equation
(\ref{eigen2}) shows that $\psi_0^+(x)$ is the ground state, which
may or may not be annihilated by $A(x)$, i.e., $A(x)\psi_0^{+}(x)
=0$, or $\neq 0$.

All eigenstates of the two partner Hamiltonians are paired when the
ground state $\psi_0^+(x)$ is not annihilated by $A(x)$. The exact
relation between the eigen-values and eigen-functions are easily
found to be %%
\begin{eqnarray}
E_n^{-} = E_{n}^{+} > 0\,,~~ \psi_n^{-}(x) = [E_{n}^{+}]^{-1/2} A(x)
\psi_{n}^{+}(x)~,~~\psi_{n}^{+}(x) = [E_{n}^{-}]^{-1/2} A^{\dag}(x)
\psi_{n}^{-}(x)~. \label{tdegeneracy1}
\end{eqnarray}
But this does not happen when the ground state $\psi_0^+(x)$ is
annihilated by $A(x)$. In this case the ground state remains
unpaired, while all other states of the two partner Hamiltonians are
paired. This time the exact relation between the eigen-values and
eigen-functions are  found to be
\begin{eqnarray}
 E_n^{-} = E_{n+1}^{+},~~ E_0^{+}  = 0~,~~\psi_n^{-}(x) = [E_{n+1}^{+}]^{-1/2}
A(x) \psi_{n+1}^{+}(x)~,~~ \psi_{n+1}^{+}(x)  = [E_{n}^{-}]^{-1/2}
A^{\dag}(x) \psi_{n}^{-}(x)~. \label{tdegeneracy2}
\end{eqnarray}
Note that when the ground state $\psi_0^+(x)$ is annihilated by
$A(x)$, the expression for the ground state can be found if the
super-potential  $W(x)$ is known. On the other hand if the ground
state is known then the expression for the super-potential can be
found from $A(x)\psi_0^{+}(x)=0$. Depending on which one is given
the ground state and the super-potential are given respectively as
follows:
\begin{eqnarray}
\psi_0^{+}(x) \sim\exp[- \sqrt{2} \int^{x} W(y) dy]\,.
\label{ground1}
\end{eqnarray}
\begin{eqnarray}
W(x) = - 2^{-\frac{1}{2}}\psi_0'^{+}(x)(\psi_0^{+}(x))^{-1}\,.
\label{ground2}
\end{eqnarray}
One can show that the above mentioned procedure can 
in fact  be repeatedly used in order to generate a
hierarchy of Hamiltonians \cite{cooper3,cooper1,cooper2,cooper4}.
For example starting with the original Hamiltonian $H_1$, which 
has $p(\geq 1)$ bound
states with eigenvalues $E^{(1)}_n$ and eigenfunctions
$\psi^{(1)}_n$ with $0\leq n \leq (p-1)$,  a hierarchy of
$(p-1)$ Hamiltonians $H_2, H_3, ... H_p$ can be constructed so 
that the m'th member of the hierarchy of Hamiltonians  $(H_m)$ have
the same eigenvalue  as $H_1$ except that the first $(m-1)$
eigenvalues of  $H_1$ are missing in  $(H_m)$.
%%%%%%%%%%%%%%%%%%%%%%%%%
\section{\small{\bf{Generalized MIC-Kepler system}}}\label{mic}
%%%%%%%%%%%%%%%%%%%%%%%%%%%%
In introduction we mentioned that generalized MIC-Kepler system
\cite{mar,mar1} is basically dynamics of two dyon  system.  The
Hamiltonian of the system is given by (in system of units $\hbar= m=
c= 1$) %%
\begin{eqnarray}
H = \frac{1}{2}\left( -i {\bf\nabla} - \mu {\bf A}\right)^2
+\frac{\mu^2}{2r^2} - \frac{1}{r} +\frac{c_1}{r^2(1+\cos{\theta})}
+\frac{c_2}{r^2(1 -\cos{\theta})}, \label{gmicKepler1}
\end{eqnarray}
Here ${\bf A}$ is the magnetic vector potential  of the Dirac
monopole, given by %%
\begin{eqnarray}
{\bf A}= -\frac{\sin\theta}{r(1-\cos\theta)}\hat\phi, \label{pot}
\end{eqnarray}
such that $curl{\bf A} = \frac{\bf r}{r^3}$. $c_1, c_2$ are
nonnegative constants and $\mu$, the Dirac quantization condition
takes the values $0, \pm \frac{1}{2},\pm 1, \cdots$.

This system has been generalized from MIC-Kepler \cite{zwa,mcl}
system by adding two axially symmetric potentials with coefficients
$c_1$ and $c_2$. It has been solved in spherical and parabolic
coordinates \cite{mar,mar1} and even generic solution \cite{giri}
has been considered by making a one parameter family of self-adjoint
extensions. Here we describe the solution in spherical polar
coordinates in order to get an idea of the generalized MIC-kepler
system, which we will need in the next section for super-symmetric
extension of the system.

The Schr\"{o}dinger eigenvalue equation is %%
\begin{eqnarray}
H\Psi = E\Psi\,, \label{eigen}
\end{eqnarray}
where the  Hamiltonian $H$ is given in Eq. (\ref{gmicKepler1}). We
can separate Eq. (\ref{eigen}) in spherical polar coordinates $(r,
\theta, \varphi)$ using the  wave-function $\Psi$  of the form %%
\begin{eqnarray}
\psi(r,\theta,\varphi)=R(r)Z(\theta, \varphi). \label{wave}
\end{eqnarray}
Substituting  Eq. (\ref{wave}) in the Schr\"{o}dinger eigenvalue
equation Eq. (\ref{eigen}) one can get  angular and radial part of
the differential equation completely separated. The angular
differential equation takes the form %%
\begin{eqnarray}
\frac{1}{\sin\theta}\frac{\partial}{\partial\theta}\left(\sin\theta\frac{\partial
Z}{\partial\theta}\right) +
\frac{1}{4\cos^2\frac{\theta}{2}}\left(\frac{\partial^{2}}{\partial\varphi^{2}}-4c_1\right)Z
+\frac{1}{4\sin^2\frac{\theta}{2}}\left[\left(\frac{\partial}{\partial\varphi}+2i\mu\right)^{2}
-4c_2\right]Z =-{\cal A}Z,\label{angular}
\end{eqnarray}
where the separation constant $\cal A$ is quantized as %%
\begin{eqnarray}
{\cal A} = \left(j+\frac{\delta_{1}+\delta_{2}}{2}\right)
\left(j+\frac{\delta_{1}+\delta_{2}}{2}+1\right). \label{seperation}
\end{eqnarray}
The solution of Eq.(\ref{angular}) can be found apart from
normalization as%%
\begin{eqnarray}
Z_{jm}^{(\mu)}(\theta, \varphi; \delta_{1}, \delta_{2} )=
\left(\cos\frac{\theta}{2}\right)^{m_1}
\left(\sin\frac{\theta}{2}\right)^{m_2}
P_{j-m_+}^{(m_2,m_1)}(\cos\theta) e^{i(m-\mu)\varphi},
\label{angularwave}
\end{eqnarray}
where $m_1=|m-\mu|+\delta_{1}=\sqrt{(m-\mu)^2+4c_1}$,
$m_2=|m+\mu|+\delta_{2}=\sqrt{(m+\mu)^2+4c_2}$,
$m_+=(|m+\mu|+|m-\mu|)/2$ and $P_n^{(a,b)}$ is the  Jacobi
polynomial. The $z$ component of the total angular momentum $m$, the
total angular momentum  $j$ take the quantized values as: %%
\begin{eqnarray}
\nonumber m &=&-j,-j+1,\dots,j-1,j\\ j &=&
\frac{|m+\mu|+|m-\mu|}{2}, \frac{|m+\mu|+|m-\mu|}{2}+1,\dots.
\end{eqnarray}
The quantum numbers $j$ and $m$ explicitly depend on the Dirac
quantization condition $\mu$. Note that depending on whether $\mu$
is half integral or integral, $j$ and $ m$ respectively  become half
integer or integer.

The radial differential equation is given by %%
\begin{eqnarray}
\frac{1}{r^{2}} \frac{d}{dr}\left(r^{2}\frac{dR}{dr}\right)-
\frac{1}{r^{2}}\left(j+\frac{\delta_{1}+\delta_{2}}{2}\right)
\left(j+\frac{\delta_{1}+\delta_{2}}{2}+1\right)R+
2\left(E+\frac{1}{r}\right)R=0\,, \label{rsolution1}\end{eqnarray}
which is exactly same as the radial equation for the hydrogen atom
except that the orbital quantum number $l$ is replaced here by
$j+(\delta_{1}+\delta_{2})/2$. The solution of  Eq.
(\ref{rsolution1}) apart from normalization takes the form %%
\begin{eqnarray}
R_{nj}^{(\mu)}(r)= (2\varepsilon r)^{j+\frac{\delta_1+
\delta_2}{2}}e^{-\varepsilon r}F\left(-n+j+1; 2j+\delta_1+
\delta_2+2; 2\varepsilon r\right)\,,
\label{rsolution2}\end{eqnarray} %%
where $n=|\mu|+1, |\mu|+2,\dots.$ and the parameter $\varepsilon$ is
given  by %%
\begin{eqnarray}
\varepsilon= \sqrt{-2E} = \frac{1}{n+\frac{\delta_1+
\delta_2}{2}}\,. \label{epsa}\end{eqnarray} %%
The exact form of the eigenvalue $E$ is now given by %%
\begin{eqnarray}
E \equiv E_n^{(\mu)} = -\frac{1}{2\left(n+\frac{\delta_1+
\delta_2}{2}\right)^{2}}\,. \label{energy}\end{eqnarray} %%
%%%%%%%%%%%%%%%%%%%%%%%%%
\section{\small{\bf{Supersymmetric generalized MIC-Kepler System}}}\label{susy}
%%%%%%%%%%%%%%%%%%%%%%%%%%%
We now come to the discussion of SUSY extension of generalized
MIC-Kepler \cite{mar,mar1} system, which is our interest in this
work.  Due to the close similarity of the radial equation
(\ref{rsolution1}) with the nonrelativistic Hydrogen atom radial
equation, we will proceed with the same line as described in
\cite{alan,alan1,alan2,alan3,alan4} for Hydrogen atom problem.

We  first remove the first order derivative term from Eq.
(\ref{rsolution1}) by the transformation $R(r)\to\chi(r)/r$. The
resulting differential equation suitable for the study of SUSY is %%
\begin{eqnarray}
-\frac{1}{2}\frac{d^2\chi(r)}{dr^2} + \left[- \frac{1}{r}- E
+\frac{\left(j+\frac{\delta_{1}+\delta_{2}}{2}\right)
\left(j+\frac{\delta_{1}+\delta_{2}}{2}+1\right)}{2r^{2}}\right]\chi(r)=0\,.
\label{radialeigen2}
\end{eqnarray}
Now we can  construct  bosonic  and fermionic one dimensional
Hamiltonian corresponding to the radial equation
(\ref{radialeigen2}). We keep $j+\frac{\delta_{1}+\delta_{2}}{2}$
fixed for our calculation of SUSY. The partner Hamiltonian $H_+(r)$
for the generalized MIC-Kepler system may be read from Eq.
(\ref{radialeigen2}) as%%
\begin{eqnarray}
H_+(r) = -\frac{1}{2}\frac{d^2}{dr^2} + V_+(r)\,,
\label{radialhamil}
\end{eqnarray}
where the SUSY partner potential  $V_+(r)$ takes the form %%
\begin{eqnarray}
V_+(r)= -\frac{1}{r} +
\frac{\left(j+\frac{\delta_{1}+\delta_{2}}{2}\right)
\left(j+\frac{\delta_{1}+\delta_{2}}{2}+1\right) }{2r^{2}} +
\frac{1}{2(j+\frac{\delta_{1}+\delta_{2}}{2}+1)^2}\,. \label{v+}
\end{eqnarray}
The potential $V_+(r)$ is constructed in such a way that the ground
state eigenvalue of the Hamiltonian Eq. (\ref{radialhamil}) becomes
zero. Because, according to the condition of  SUSY in Eq,
(\ref{tdegeneracy2}), we need the ground state eigenvalue of the
partner Hamiltonian $H_+$ to be zero. The constant term
$\frac{1}{2(j+\frac{\delta_{1}+\delta_{2}}{2}+1)^2}$ in Eq.
(\ref{v+}) guarantees that the ground state eigenvalue of the
Hamiltonian in Eq. (\ref{radialhamil}) is zero for fixed  angular
momentum quantum number $j$. The expression for SUSY partner
potential $V_+(r)$ in  Eq. (\ref{v+}) exactly matches with the SUSY
partner potential $V_+(r)$ of the Hydrogenic problem if
$j+\frac{\delta_{1}+\delta_{2}}{2}$ in Eq. (\ref{v+}) is replaced by
the orbital angular momentum quantum number $l$
\cite{alan,alan1,alan2,alan3,alan4}.  The energy eigenvalue of the
partner Hamiltonian $H_+(r)$ Eq. (\ref{radialhamil}) takes the form
\begin{eqnarray}
E_n^+= \frac{1}{2(j+\frac{\delta_{1}+\delta_{2}}{2}+1)^2} -
\frac{1}{2(n
  +\frac{\delta_{1}+\delta_{2}}{2})^2}~~~ \mbox{for},~~ n \geq j+1\,.
\label{spectrum1}
\end{eqnarray}

Note that the energy eigenvalue $E_n^+$ of the partner Hamiltonian
$H_+(r)$ does not reduce to the energy eigenvalue of the
corresponding  partner Hamiltonian  of the Hydrogenic problem after
replacement of $j+\frac{\delta_{1}+\delta_{2}}{2}$ by orbital
angular momentum quantum number $l$ in Eq. (\ref{spectrum1}) unlike
SUSY partner potential $V_+(r)$. However for $\mu= c_1= c_2 = 0$,
$E_n^+$ of Eq (\ref{spectrum1}) reduces to the corresponding
eigenvalue of the Hydrogenic problem and the same is applicable for
SUSY partner potential $V_+(r)$. From Eq. (\ref{partnerpotential1})
and Eq. (\ref{v+}) we can work out the super-potential $W(r)$ as %%
\begin{eqnarray}
W(r)= \frac{1}{\sqrt{2}(j+\frac{\delta_{1}+\delta_{2}}{2}+1)} -
\frac{j+\frac{\delta_{1}+\delta_{2}}{2}+1}{\sqrt{2}r}\,.
\label{superpotential}
\end{eqnarray}
One may note here also that the super-potential $W(r)$ exactly
matches with the super-potential of the Hydrogenic problem
\cite{alan,alan1,alan2,alan3,alan4} if
$j+\frac{\delta_{1}+\delta_{2}}{2}$ in Eq.  (\ref{superpotential})
is replaced by $l$. From Eq.  (\ref{partnerpotential1}) and Eq.
(\ref{superpotential}) we can calculate the  SUSY partner potential
$V_-$ as %%
\begin{eqnarray}
V_-(r)=  -\frac{1}{r} +
\frac{\left(j+\frac{\delta_{1}+\delta_{2}}{2} +1\right)
\left(j+\frac{\delta_{1}+\delta_{2}}{2}+2\right) }{2r^{2}} +
\frac{1}{2(j+\frac{\delta_{1}+\delta_{2}}{2}+1)^2}\,. \label{v-}
\end{eqnarray}
It is now obvious from the above discussion of super-potential
$W(r)$ that we may get partner potential of the Hydrogenic problem
from Eq.  (\ref{v-}) by the same replacement.  Similarly we  may get
the eigenvalue $E_n^-$  of the SUSY partner Hamiltonian $H_-(r)$
from the relation Eq. (\ref{tdegeneracy2}) and Eq.
(\ref{spectrum1}).  From the above analysis one can make  the
following important observations in addition to the non-relativistic
coulomb problem discussed in \cite{alan,alan1,alan2,alan3,alan4}.

%%%%%%%%%%%%%%%%%%%%%%%%%
\subsection{\small{\bf{Observation 1}}}
%%%%%%%%%%%%%%%%%%%%%%%%%
Consider the situation  $c_1= c_2 = \mu= 0$. In this case the
Hamiltonian in Eq. (\ref{gmicKepler1}) reduces to the Hydrogen atom
problem. Putting the above condition in Eq. (\ref{v+}) and Eq.
(\ref{v-}) we can immediately get the SUSY partner  potentials
$V_{hydrogen +}(r)$  and  $V_{hydrogen-}(r)$ respectively  for the
Hydrogen atom problem  as %%
\begin{eqnarray}
V_{hydrogen +}(r)&=& -\frac{1}{r} + \frac{l(l+1)}{2r^{2}} +
\frac{1}{2(l+1)^2}\,, \label{v+o1}\\ V_{hydrogen -}(r)&=&
-\frac{1}{r} + \frac{(l+1)(l+2)}{2r^{2}} + \frac{1}{2(l+1)^2}\,,
\label{v+o12}
\end{eqnarray}
and we may also get the eigenvalues $(E_{hydrogen})_ n^+$ and
$(E_{hydrogen})_ n^-$  for the Hydrogen atom problem, putting the
same condition in Eq. (\ref{spectrum1})  as %%
\begin{eqnarray}
(E_{hydrogen})_n^+& =& \frac{1}{2(l+1)^2} - \frac{1}{2(n)^2}~~~
\mbox{for},~~ n \geq l+1 \label{spectrumo1}\\ (E_{hydrogen})_{n}^-&
=& (E_{hydrogen})_{n+1}^+~~~ \mbox{for},~~ n \geq l+1\,.
\label{spectrumo12}
\end{eqnarray}
However  the results  Eq. (\ref{v+o1}) - Eq. (\ref{spectrumo12})
are not new
 to us. It has already  been calculated in Ref.
\cite{alan,alan1,alan2,alan3,alan4}. Although
 the physical meaning of these results was explained  in
\cite{alan,alan1,alan2,alan3,alan4};
 here we briefly discuss about  that. Consider the case $l=0$ ($s$
 orbitals). We can see from Eq. (\ref{spectrumo1}) and
 Eq. (\ref{spectrumo12}) that  the spectrum of $H_-$  (for $l=1$ ) is same as that of $H_+$
 ( for $l=0$ )  with the ground state removed. So $H_-$ ( for $l=1$ ) may
 describe a system with all the $s$ orbitals removed. It can have the
 following  physical description. For lithium atom we may have this kind of
 situation where two of the three electrons of the lithium atom are in the
 $1s$ orbital.  So for lithium atom if the electron-electron interaction is
 taken to be zero and if the valence electron is far away moved from the core
 electrons then the effective interaction will look like  Coulombic.
 Thus the $l=0$ levels of the lithium atom
 becomes the SUSY partner of the Hydrogen atom $l=0$ levels.
%%%%%%%%%%%%%%%%%%%%%%%%%
\subsection{\small{\bf{Observation 2}}}
%%%%%%%%%%%%%%%%%%%%%%%%%
Consider the situation  $c_1= c_2 =  0$ of the Hamiltonian in
 Eq. (\ref{gmicKepler1}). This is the two particle problem which have magnetic
 charges $g_1, g_2$ in addition to their electric charges  $e_1,e_2$.  The
 potential $V_\pm(r)$ take the form
\begin{eqnarray}
V_+(r) &=& -\frac{1}{r} + \frac{j \left(j +1\right) }{2r^{2}} +
\frac{1}{2(j+1)^2}\,,
\label{v+o2}\\
V_-(r) &=& -\frac{1}{r} + \frac{(j+1) \left(j +2\right) }{2r^{2}} +
\frac{1}{2(j+1)^2}\,, \label{v+o21}
\end{eqnarray}
and the corresponding eigenvalues are given by %%
\begin{eqnarray}
E_n^+ &=& \frac{1}{2(j+1)^2} - \frac{1}{2(n)^2}~~~ \mbox{for},~~ n
\geq j+1
\label{spectrumo2}\\
E_n^-&=&E_{n+1}^+~~~ \mbox{for}, ~~ n \geq j+1\,.
\label{spectrumo21}
\end{eqnarray}
It is to be noted that the expressions in Eq. (\ref{v+o2}) - Eq.
(\ref{spectrumo21}) look like Hydrogenic problem in Eq. (\ref{v+o1})
- Eq. (\ref{spectrumo12}), but it is not the same problem because
$j$ can take values $j = |\mu|, |\mu| +1, |\mu| +2,.....$, whereas
for Hydrogenic problem $l$ can take non negative integral  values
only.  If one considers the case of lowest value of $j$, i.e., $j=
|\mu|$, then it is easy to show that the systems with Dirac
quantization condition differing by $|\mu_1|- |\mu_1|= 1$ are
super-symmetric partners. Since $\mu$ can take values like $\mu= 0,
\pm \frac{1}{2}, \pm 1, \pm \frac{3}{2}, ....$, it is
straightforward to conclude that system with $\mu= \pm \frac{1}{2},
\pm\frac{3}{2},....$ belong to one SUSY family and system with $\mu=
0, \pm 1, \pm 2,....$ belong to other SUSY family. The two family
gets decoupled in this scenario. In the next subsection we will show
that it is necessary to incorporate the additional potentials with
the MIC-Kepler system, as it has been done in Refs. \cite{mar,mar1},
to get a unified SUSY family.
%%%%%%%%%%%%%%%%%%%%%%%%%
\subsection{\small{\bf{Observation 3}}}
%%%%%%%%%%%%%%%%%%%%%%%%%
Finally, consider the situation when all the three parameters $\mu,
c_1, c_2$ are nonzeros.  The potential $V_+(r)$ and the respective
energy eigenvalue $E^+_n$ is given by Eq. (\ref{v+}) and Eq.
(\ref{spectrum1}) respectively. Now consider the situation when
$\delta_1+\delta_2=1$. This can be achieved by appropriately  tuning
the constant parameter $c_1$ and $c_2$ in the Hamiltonian
(\ref{gmicKepler1}). Replacing $\delta_1+\delta_2=1$ in Eq.
(\ref{v+}) and Eq. (\ref{spectrum1}), we can see that systems with
half integral Dirac quantization condition,  $\mu= \pm \frac{1}{2},
\pm\frac{3}{2},....$, belong to the family of integral Dirac
quantization condition.

%%%%%%%%%%%%%%%%%%%%%%%%%
\section{\small{\bf{Conclusion}}}\label{con}
%%%%%%%%%%%%%%%%%%%%%%%%%
SUSY has been shown to be a good symmetry for nonrelativistic
coulomb system \cite{alan,alan1,alan2,alan3,alan4}. It has been
successfully applied to many other systems also, for example, in
harmonic oscillator system \cite{cooper3} it has been applied to get
supersymmetric system. We however show in our calculation that if
SUSY is conserved in  MIC-Kepler system, then there exist two kind
of super-symmetric families, one with half integral Dirac
quantization condition   $\mu= \pm \frac{1}{2}, \pm\frac{3}{2},....$
and other with integral Dirac quantization condition $\mu= 0, \pm 1,
\pm 2,....$. We also show that in order to unify these two kind of
SUSY families into one, we need to generalize MIC-Kepler system
\cite{mar,mar1}. Once we generalize, the two extra potentials with
coefficients $c_1$ and $c_2$ in the Hamiltonian (\ref{gmicKepler1})
allows us to unify the apparently separated SUSY families. We
reproduce the non-relativistic coulomb result
\cite{alan,alan1,alan2,alan3,alan4} in our formulation in subsection
(4.1). The physics behind this SUSY can be understood from the SUSY
problem of nonrelativistic Coulomb problem discussed in subsection
(4.1). Since nonrelativistic Coulomb problem is a special case of
Generalized MIC-Kepler system, one can expect similar physics for
Generalized MIC-Kepler system also. But in order to see this
symmetry in nature, we first need dyon to exist in nature. SUSY
breaking and self-adjointness is also an interesting issue
\cite{pisani}. We hope to discuss this for generalized MIC-Kepler
system in future.
%%%%%%%%%%%%%%%%%%%%%%%%%%%
\subsubsection*{Acknowledgments}
%%%%%%%%%%%%%%%%%%%%%%%%%%%
We thank  Palash B. Pal for comments on manuscript and helpful
discussions.
%%%%%%%%%%%%%%%%%%%%%%%%%%%

\end{document}